*General Computational Biology*

# Uncovering the dynamic effects of DEX treatment on lung cancer by integrating bioinformatic inference and multiscale modeling of scRNA-seq and proteomics data

Minghan Chen[1], Chunrui Xu[2], Ziang Xu[1,3], Wei He[2], Haorui Zhang[4], Jing Su[5], and Qianqian Song[6,7*]

[1]Department of Computer Science, Wake Forest University, Winston-Salem, NC, USA

[2]Genetics, Bioinformatics, and Computational Biology, Virginia Tech, Blacksburg, VA, USA

[3]Department of Chemistry, Wake Forest University, Winston-Salem, NC, USA

[4]Department of Mathematics and Statistics, Wake Forest University, Winston-Salem, NC, USA

[5]Department of Biostatistics and Health Data Science, Indiana University School of Medicine, Indianapolis, IN, USA

[6]Center for Cancer Genomics and Precision Oncology, Wake Forest Baptist Comprehensive Cancer Center, Wake Forest Baptist Medical Center, Winston Salem, NC, USA

[7]Department of Cancer Biology, Wake Forest School of Medicine, Winston Salem, NC, USA

*To whom correspondence should be addressed.

## Abstract

**Motivation:** Lung cancer is one of the leading causes for cancer-related death, with a five-year survival rate of 18%. It is a priority for us to understand the underlying mechanisms that affect the implementation and effectiveness of lung cancer therapeutics. In this study, we combine the power of Bioinformatics and Systems Biology to comprehensively uncover functional and signaling pathways of drug treatment using bioinformatics inference and multiscale modeling of both scRNA-seq data and proteomics data. The innovative and cross-disciplinary approach can be further applied to other computational studies in tumorigenesis and oncotherapy.

**Results:** A time series of lung adenocarcinoma-derived A549 cells after DEX treatment were analysed. **(1)** We first discovered the differentially expressed genes in those lung cancer cells. Then through the interrogation of their regulatory network, we identified key hub genes including TGF-β, MYC, and SMAD3 varied underlie DEX treatment. Further enrichment analysis revealed the TGF-β signaling pathway as the top enriched term. Those genes involved in the TGF-β pathway and their crosstalk with the ERBB pathway presented a strong survival prognosis in clinical lung cancer samples. **(2)** Based on biological validation and further curation, a multiscale model of tumor regulation centered on both TGF-β-induced and ERBB-amplified signaling pathways was developed to characterize the dynamics effects of DEX therapy on lung cancer cells. Our simulation results were well matched to available data of SMAD2, FOXO3, TGFβ1, and TGFβR1 over the time course. Moreover, we provided predictions of different doses to illustrate the trend and therapeutic potential of DEX treatment.

**Availability:** https://github.com/chenm19/DEXCancer

**Contact:** qsong@wakehealth.edu

**Supplementary information:** Supplementary data are available at *Bioinformatics* online.

## 1 Introduction

Lung cancer remains the most common cause of cancer-related death, which occupies 18% of mortality worldwide for all 36 cancers and with over 1.8 million deaths expected globally in 2021. The high lung cancer fatality rate is primarily attributed to the large proportion of patients (57%) diagnosed with metastatic disease, for which the five-year relative survival rate is 18% (Siegel, et al., 2021). Therefore, effective therapeutic strategies are in urgent need. Glucocorticoids (GCs), which are commonly used to treat autoimmune disorders, are continually investigated as a potential strategy since they can be employed to treat inflammation as well as to enhance the anti-tumor effect of drugs in chemotherapy (Herr and Pfitzenmaier, 2006; Wang, et al., 2007). As one of the most widely used synthetic glucocorticoids, Dexamethasone (DEX) has shown anti-cancer efficacy and anti-estrogenic activity in human non-small cell lung cancer (NSCLC) whereby further study can be made to testify its efficacy.

Single-cell RNA sequencing (scRNA-seq) provides unprecedented opportunities for understanding cellular complexity (Athanasiadis, et al., 2017; Chu, et al., 2016; Macaulay, et al., 2016; Patel, et al., 2014; Shin, et al., 2015). Transcriptomic profiling in individual cells has revealed a variety of cell types and subpopulations. scRNA-seq not only provides profound insights into cellular composition, but also allows for the interrogation of cellular hierarchies and the identification of cells transitioning between states (Baron, et al., 2016; Jia, et al., 2018; Liu, et al., 2017), such as development and differentiation. Single-cell RNA sequencing is at the forefront of phenotyping cells with high resolution and has been widely applied in research and revealed the molecular determinants of human diseases (Macaulay, et al., 2016). For example, Welch J. et.al. revealed the putative mechanisms of cell-type-specific regulation within their defined mouse cortical cell types (Welch, et al., 2019). Nativio R. et.al (Nativio, et al., 2020) identified molecular pathways underlying late-onset Alzheimer's disease. Bian S. et.al. (Bian, et al., 2018) reconstructed genetic lineages and traced the transcriptomic dynamics through single-cell profiling. These studies highlight the significance of single-cell technology in accelerating the investigations of gene regulation and illuminating the causes and underlying mechanisms of human diseases especially cancers (Fu, et al., 2021).

However, bioinformatics approaches rely on scRNA-seq data to infer gene regulatory networks, which only address "mechanisms" by proxy (i.e., correlations) without providing the underlying information (i.e., how genes interact with one another via activation, inhibition, or binding pathways). Mathematical modeling based on experiments is a central approach of systems biology, which has been applied to investigate complex signaling transduction pathways of tumorigenesis (Kirouac, et al., 2013; Labibi, et al., 2020; Wang, et al., 2009). Indeed, modeling signaling transduction and cross-talk gives insights into not only the regulatory networks of oncogenes but also the effects of anti-tumor treatments. The transforming growth factor β (TGF-β) –induced SMAD signaling is a quintessential pathway to regulate cancer progression (Kang, et al., 2005), which is one core study object of modeling (Zi, et al., 2011). Additionally, systems biologists have investigated cross-talks and feedback loops among tumor signaling transduction pathways to understand tumorigenesis and treatments, such as cross-talks between the TGF-β-induced pathway and ERBB-amplified proliferation pathway, and the negative feedbacks within the tumor growth and proliferation pathways (Kirouac, et al., 2013).

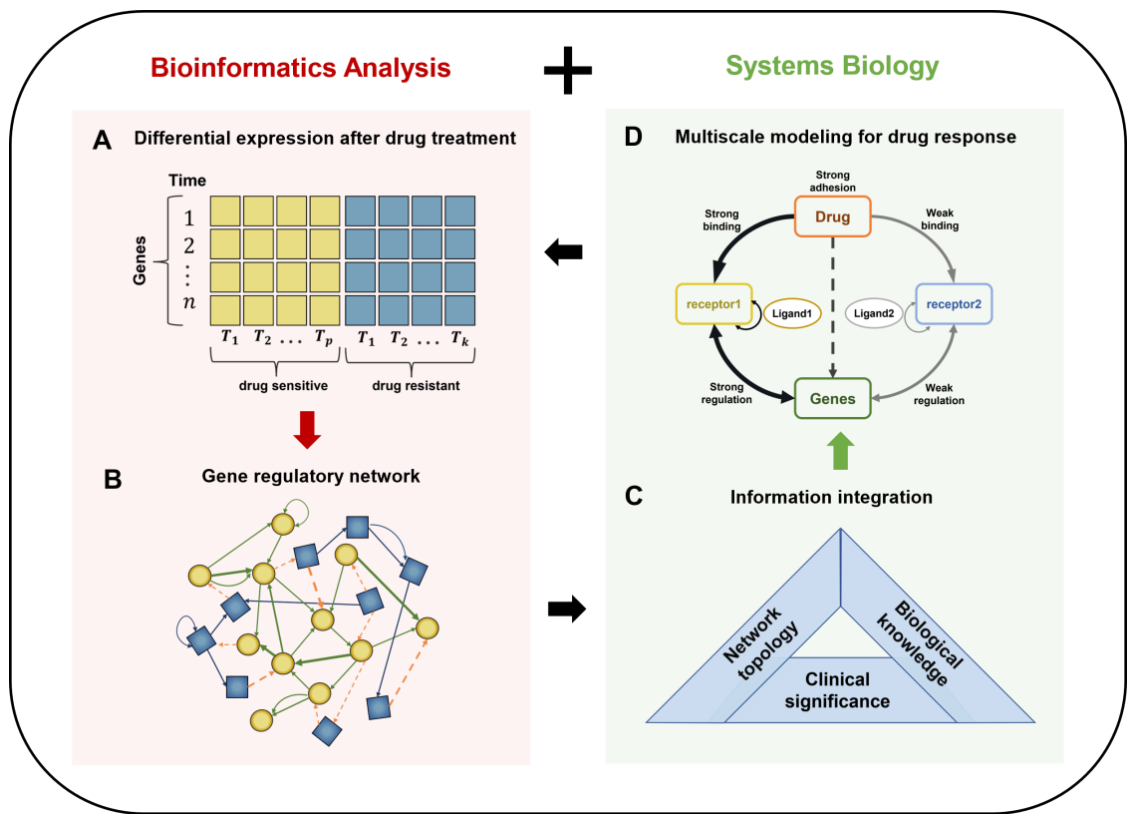

**Fig. 1. Schematic illustration of integrated Bioinformatics (anatomizing, reconstruction) and Systems Biology (integration, modeling) analysis on drug effect. (A)** Analyze the scRNA-seq dataset of lung adenocarcinoma derived A549 cells with dexamethasone (DEX) treatment, then categorize genes into two types based on the pharmacokinetic properties of either drug sensitive or drug resistant. **(B)** Ground on the cell type-specific gene expression, then construct the gene regulatory multilayer network with crosstalk and feedback loops. **(C)** The important genes from network topology, biological knowledge, and clinical significance are identified and integrated for further modeling. **(D)** Develop a multiscale model of interactive genes, ligand, receptors, and drug to characterize drug response.

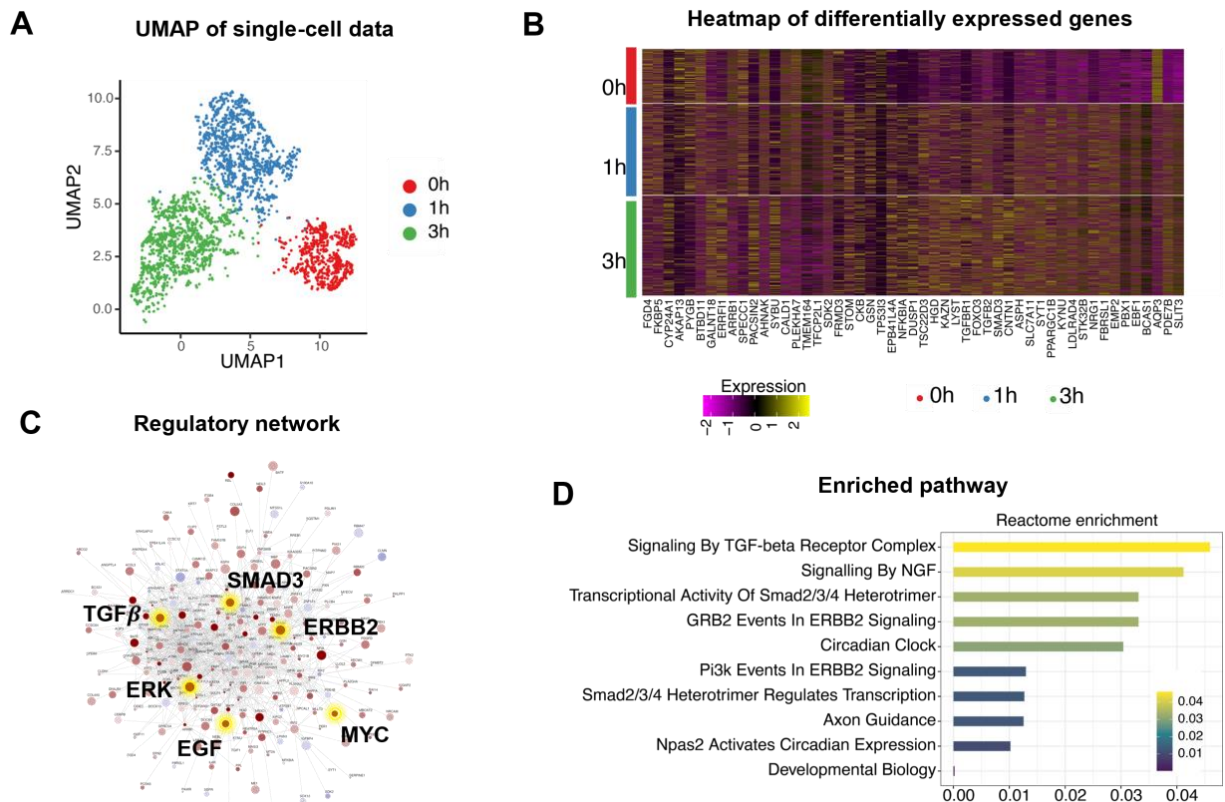

**Fig. 2. Inference of signaling network in lung cancer cells after drug treatment.** (**A**) UMAP visualization of A549 cells, colored by DEX treatment time (0h, 1h, and 3 h). (**B**) Heatmap visualization of differentially expressed genes (DEGs) identified under different treatment time. (**C**) Transcriptional regulatory network inferred based on the DEGs after DEX treatment. (**D**) Enriched pathways identified based on the transcriptional regulatory network.

In this study, we will integrate the knowledge and approaches from both Bioinformatics and Systems Biology domains to uncover the underlying molecular mechanisms of DEX therapy on lung cancer cells. With the collected lung adenocarcinoma-derived A549 cells after DEX treatment, we will explore key differentially expressed genes and their involved regulation pathways after DEX. Furthermore, we develop a multiscale model to investigate the underlying mechanisms of anti-tumor drugs and examine the regulatory networks obtained by the above bioinformatical analysis. The systematic approach based on single-cell transcriptomics data and proteomics data can better delineate the dynamic and longitudinal signaling pathways mediated by DEX in lung cancer cells.

## 2   Methods

A schematic illustration of our work is shown in Figure 1, including the identification of signaling pathways, kinetic modeling, and drug prediction. We first analyzed the scRNA-seq data to identify the variable genes and the involved interaction network. The network involved genes were used to infer intracellular signaling pathways associated with clinical significance. We further built the mathematical model based on the signaling pathways to mechanistically understand the dynamic regulations of drug response involved in lung cancer cells.

**Bioinformatics methods.** Uniform manifold approximation and projection (UMAP) was used to visualize cell clusters. Differential expressed genes (DEGs) were identified using the FindMarkers function in Seurat (Butler, et al., 2018). To identify putative transcriptional regulators of the DEGs, we utilized the GENIE (Huynh-Thu, et al., 2010) method. Significant regulatory networks associated with the upregulated and downregulated genes were identified.

**Survival analysis.** Kaplan-Meier (KM) analysis was performed using the "survival" R package (http://cran.r-project.org/web/packages/survival/index.html). Log-rank test was used to test the differences of survival curves.

**Pathway database.** Reactome (http://www.reactome.org) was a manually curated open-data resource of human pathways and reactions, which was an archive of biological processes and a tool for discovering potential functions. Gene sets derived from the Reactome (Fabregat, et al., 2018) pathway database was downloaded from the MSigDB Collections.

**Enrichment test.** Functional enrichment based on the Reactome and Gene Ontology (GO) databases was assessed by hypergeometric test, which was used to identify a priori-defined gene set that showed statistically significant differences between two given clusters. Enrichment test was performed by the clusterProfiler package (Yu, et al., 2012). Test P values were further adjusted by Benjamini-Hochberg correction, and adjusted P values less than 0.05 were considered statistically significant.

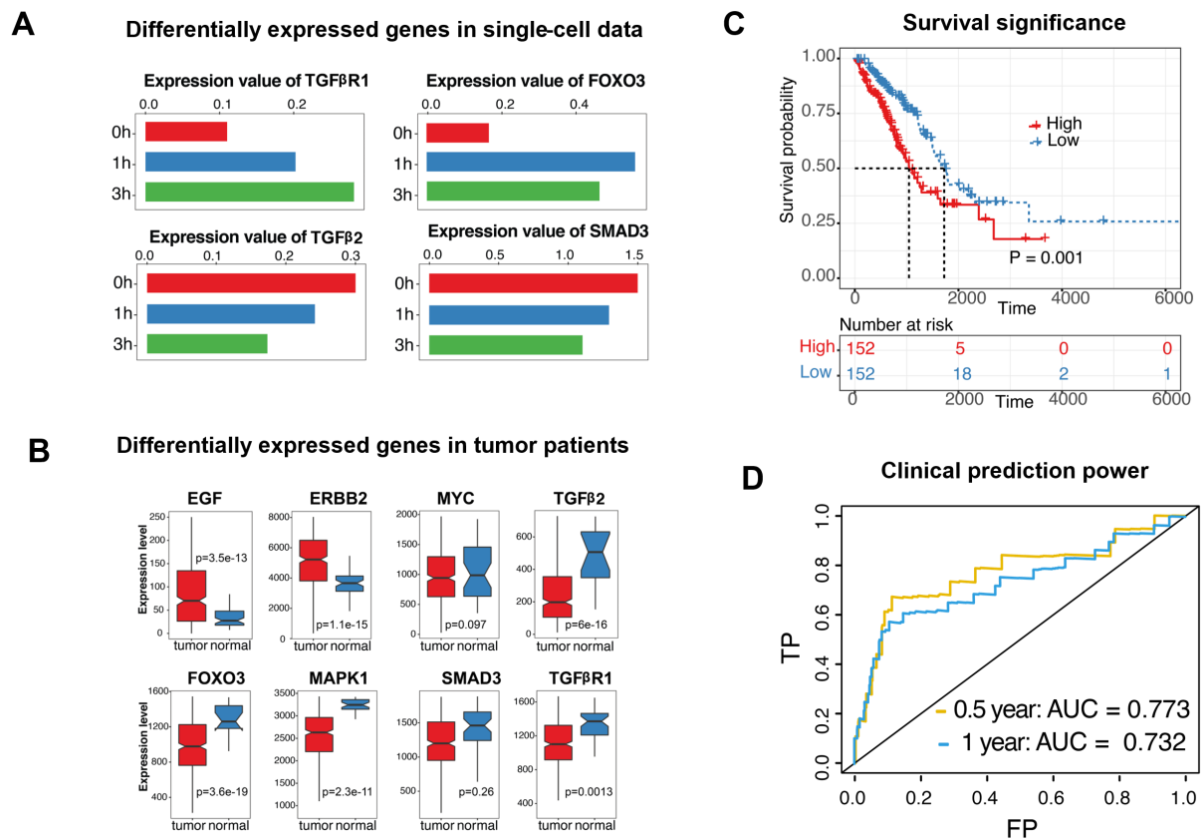

**Fig. 3. Clinical prognosis and significance of network hub genes.** (**A**) Varied expression of genes involved in both transcriptional regulatory network and enriched pathways. (**B**) Boxplots illustrated the differential expressions of selected genes between tumor and normal tissues. (**C**) Kaplan-Meier survival curves for patients with LUAD, which were stratified (binary: high versus low) for the average expression of selected genes. Log-rank test P-values were shown. The y-axis was the probability of overall survival, and the x axis was time in days. (**D**) Prognostic accuracy of the selected genes was evaluated by AUC of the time-dependent ROC with respect to half year and 1 year survival of LUAD patients in the TCGA dataset.

**Modeling methods.** Integrating genes identified based on scRNA-seq data (Figure 1B), relevant experimental knowledge (Figure 1C), and hypotheses, we constructed an overall wiring diagram to depict relationships between genes and proteins (Figure 1D). In this study, the diagram was translated into a system of ordinary differential equations (ODEs) based on biochemical rationales to describe reactions of synthesis, degradation, phosphorylation, dephosphorylation, and so on. The simulated dynamics of variables are visualized after parameterization and evaluation using available data.

**Data availability.** The time-series scRNA-seq dataset of lung adenocarcinoma derived A549 cells with dexamethasone (DEX) treatment is profiled using the sci-CAR protocol (Cao, et al., 2018), which consists of cells after 0, 1, or 3 hours of 100 nM DEX treatment. The scRNA-seq can be downloaded from GEO with accession number of GSM3271040. After quality control, the resulted data consists of 2,641 cells and 1,185 genes in our analysis. The number of cells after 0, 1 and 3 hours of 100 nM DEX treatment are 583, 983, 1075 respectively.

# 3 Results

## 3.1 Bioinformatic inference of key signaling pathways

The scRNA-seq data was shown with the UMAP projection (Figure 2A), with clearly discerned cell populations (0, 1, and 3 h cells after 100 nM dexamethasone (DEX) treatment. With the scRNA-seq data, we identified the differentially expressed genes (DEGs) for each of the cell populations, which were shown in the heatmap (Figure 2B). Those DEGs with adjusted P-value <0.01 were listed in respective of each cell states (0, 1, and 3 h), which provided intracellular clues underlie lung cancer cells after DEX treatment. To identify the signaling pathways and key regulators, we utilized the GENIE3 method to construct a regulatory network based on the DEGs (Figure 2C). Notably, genes including TGFβR1, SMAD3, ERBB2, ERK, EGF, MYC were observed as the transcriptional hub, suggesting their key roles underlie drug response. Enrichment analysis using the REACTOME pathway database (Figure 2D) were shown in bar plots. The DEGs involved in the transcriptional network were significantly enriched in TGF-beta receptor complex and ERBB2 signaling pathway, highlighting the underlying functional mechanisms. These biological functions suggest the crosstalk of TGF-β signaling pathway and ERBB signaling pathway underlie cancer cells after drug treatment.

From the scRNA-seq data, the network hub genes showed strong variations in lung cancer cells during different time points after DEX treatment (Figure 3A). To further characterize these genes, we collected the TCGA lung adenocarcinoma (LUAD) data, and performed differential expression analysis between tumor and normal samples using DEseq2(Love, et al., 2014). Interestingly, the network hub genes, such as EGF and ERBB2 were shown with higher expression levels in tumor samples (Figure 3B). There were no significant differences of MYC and SMAD3 between tumor and normal samples. In contrast, genes like FOXO3 were observed with lower expression levels in tumor than normal samples.

Additionally, we evaluated whether those network hub genes would impact patient survival. Remarkably, significant associations were observed between their increased expression and decreased overall survival of LUAD patients from TCGA (Figure 3C). The statistical significance of the difference between the K-M survival curves for patients in the high-risk group (blue) and low-risk group (red) was assessed using the log-rank test, with P-values as 0.001. The high-risk group of patients had shorter overall survival time than the low-risk group. This result suggested the prognostic roles of those network hub genes in lung cancer patients. Moreover, we also evaluated the prognostic accuracy of the network hub genes by calculating the AUC of time-dependent ROC with respect to half year and one year survival of lung cancer patients. The hub genes showed good prognostic accuracy based on the TCGA LUAD dataset (Figure 3D), demonstrating that these genes possessed convincingly strong prognostic power in predicting overall survival rates of lung cancer patients.

## 3.2 Multiscale modeling of tumorigenesis regulatory network

Based on the tumor-relevant genes identified in bioinformatic approaches (Figure 2C, 3C), we construct a tumorigenesis regulatory network summarized in Figure 4, composed of the TGF-β-induced tumor suppression pathway and ERBB-amplified proliferation pathway.

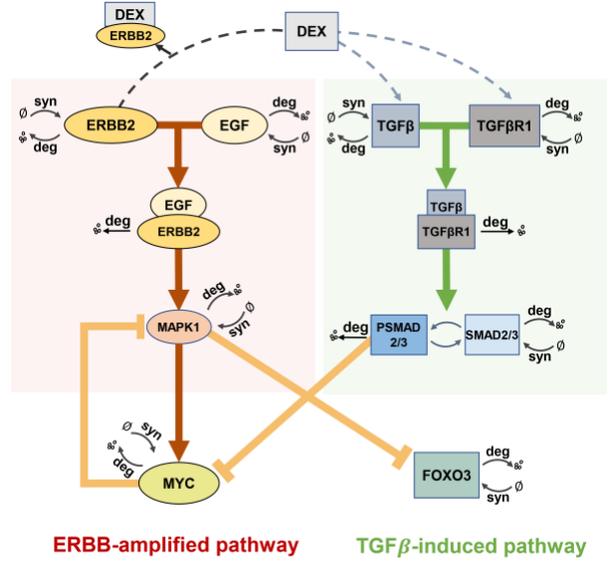

**Fig. 4. Multiscale modeling of the ERBB-amplified and TGF-β-induced feedback-crosstalk module that depicts the dynamic interplay between multilayer signals in signature genes.** The ligand EGF (epidermal growth factor) bind to the cell-surface receptors ERBB2, which activates the expression of MAPK1, thereby promoting tumor growth. The ligand TGFβ (transforming growth factor) bind to the cell-surface receptors TGFβ-R1, and their complex activates the phosphorylation of SMAD2/3. MAPK1 represses transcription factors MYC and FOXO3. Concurrently, SMAD2/3 inhibits MYC, which represses the circuit of MAPK1 mutually. Solid lines denote activation/inhibition of the shown genes with arrow/bar, respectively. The dashed lines denote the potential binding between DEX and ERBB2 and activation on TGFβ and TGFβ-R1.

**Table 1.** Equations of ERBB-amplified, TGF-β-induced, and DEX-related pathways

| | **Equations of ERBB-amplified proliferation pathway** |
|---|---|
| (1) | $dy[\text{EGF}]/dt = ks_{ef} - kd_{ef} * [\text{EGF}] - kb_{efeb} * [\text{EGF}] * [\text{ERBB2}] + ku_{efeb} * [\text{EGF\_ERBB2}]$ |
| (2) | $dy[\text{ERBB2}]/dt = ks_{eb} - kd_{eb} * [\text{ERBB2}] - kb_{efeb} * [\text{EGF}] * [\text{ERBB2}] + ku_{efeb} * [\text{EGF\_ERBB2}]$ $- kb_{dgeb} * [\text{DRUG}] * [\text{ERBB2}] + ku_{dgeb} * [\text{ERBB2\_DRUG}]$ |
| (3) | $dy[\text{EGF\_ERBB2}]/dt = -kd_{efeb} * [\text{EGF\_ERBB2}] + kb_{efeb} * [\text{EGF}] * [\text{ERBB2}] - ku_{efeb} * [\text{EGF\_ERBB2}]$ |
| (4) | $dy[\text{MAPK1}]/dt = ks_{mk} * \left(1 + \alpha_{efeb} * [\text{EGF\_ERBB2}] + \beta_{mcmk} / (km_{mcmk} + [\text{MYC}])\right) - kd_{mk} * [\text{MAPK1}]$ |
| (5) | $dy[\text{FOXO3}]/dt = ks_{fo} * (1 + \beta_{mkfo} / (km_{mkfo} + [\text{MAPK1}])) - kd_{fo} * [\text{FOXO3}]$ |
| | **Equations of TGF-β-induced pathway** |
| (6) | $dy[\text{TGF}\beta2]/dt = ks_{tb} * \left(1 + \alpha_{tbdg} * [\text{DRUG}]\right) - kd_{tb} * [\text{TGF}\beta2] - kb_{tbtr} * [\text{TGF}\beta2] * [\text{TGF}\beta\text{R1}]$ $+ ku_{tbtr} * [\text{TGF}\beta2\_\text{TGF}\beta\text{R1}]$ |
| (7) | $dy[\text{TGF}\beta\text{R1}]/dt = ks_{tr} * \left(1 + \alpha_{trdg} * [\text{DRUG}]\right) - kd_{tr} * [\text{TGF}\beta\text{R1}] - kb_{tbtr} * [\text{TGF}\beta2] * [\text{TGF}\beta\text{R1}]$ $+ ku_{tbtr} * [\text{TGF}\beta2\_\text{TGF}\beta\text{R1}]$ |
| (8) | $dy[\text{TGF}\beta2\_\text{TGF}\beta\text{R1}]/dt = -kd_{tbtr} * [\text{TGF}\beta2\_\text{TGF}\beta\text{R1}] + kb_{tbtr} * [\text{TGF}\beta2] * [\text{TGF}\beta\text{R1}] - ku_{tbtr} * [\text{TGF}\beta2\_\text{TGF}\beta\text{R1}]$ |
| (9) | $dy[\text{SMAD2/3}]/dt = ks_{sd} - kd_{sd} * [\text{SMAD2/3}] - kpho_{sd} * (1 + \alpha_{tbtr} * [\text{TGF}\beta2\_\text{TGF}\beta\text{R1}]) * [\text{SMAD2/3}]$ $+ kdepho_{sd} * [\text{pSMAD2/3}]$ |
| (10) | $dy[\text{pSMAD2/3}]/dt = -kd_{sd} * [\text{pSMAD2/3}] + kpho_{sd} * (1 + \alpha_{tbtr} * [\text{TGF}\beta2\_\text{TGF}\beta\text{R1}]) * [\text{SMAD2/3}]$ $- kdepho_{sd} * [\text{pSMAD2/3}]$ |
| (11) | $dy[\text{MYC}]/dt = ks_{mc} * (1 + \alpha_{mkmc} * [\text{MAPK1}] + \beta_{sdmc} / (km_{psdmc} + [\text{pSMAD2/3}])) - kd_{mc} * [\text{MYC}]$ |
| | **Equations of drug-related pathway** |
| (12) | $dy[\text{ERBB2\_DRUG}] = -kd_{dg} * [\text{ERBB2\_DRUG}] + kb_{dgeb} * [\text{DRUG}] * [\text{ERBB2}] - ku_{dgeb} * [\text{ERBB2\_DRUG}]$ |
| (13) | $dy[\text{DRUG}] = -kd_{dg} * [\text{DRUG}] - kb_{dgtr} * [\text{DRUG}] * [\text{TGF}\beta\text{R1}] + ku_{dgtr} * [\text{TGF}\beta\text{R1\_DRUG}]$ $- kb_{dgeb} * [\text{DRUG}] * [\text{ERBB2}] + ku_{dgeb} * [\text{ERBB2\_DRUG}]$ |

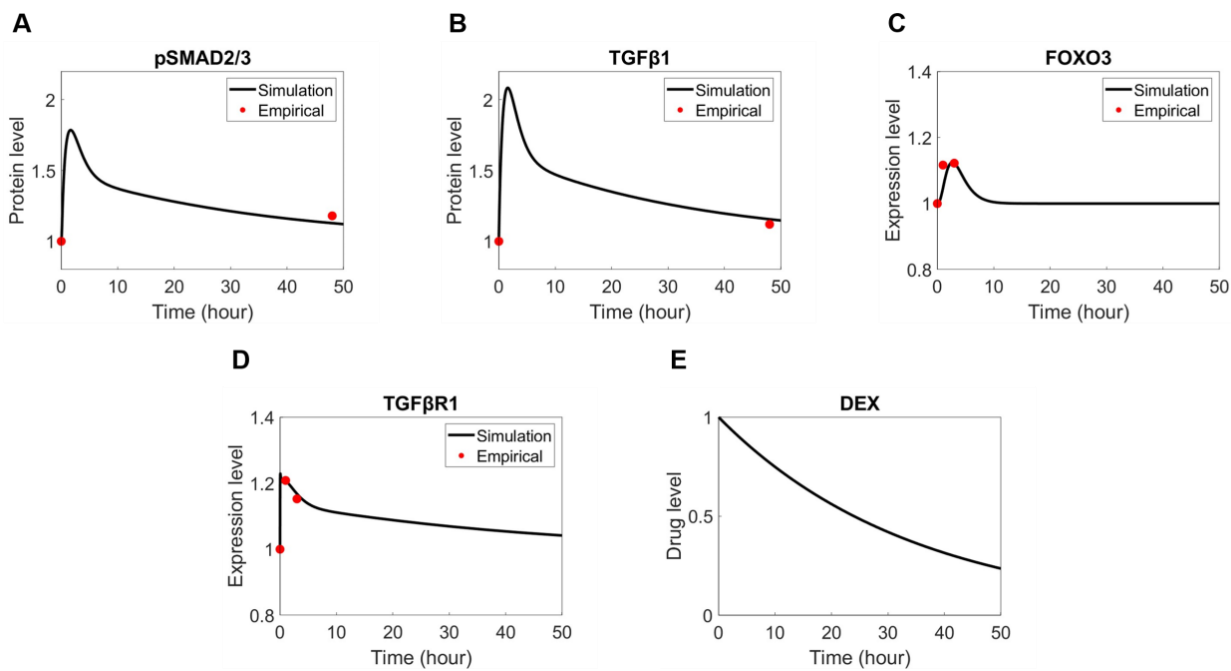

**Fig. 5.** (**A-B**) Simulated protein profiles of signature genes SMAD2 and TGFβ1after employing one dose of DEX treatment with corresponding empirical data (red circles) obtained from A549 cells after 0 and 48 hours of 100 nM DEX treatment for protein profiles (1.18 and 1.12 respectively at 48 hour) (Feng, et al., 2018). (**C-D**) Simulated expression level of signature genes FOXO3 and TGFβR1 after 0, 1, and 3 hours of 100 nM DEX treatment for gene expression levels (one dose). (**E**) DEX level over time after one dose of DEX treatment applied to the A549 cells. Note that all expression and protein levels are scaled to one at $t = 0$.

The ERBB family is one quintessence of tumor development regulation, which controls multiple downstream signaling pathways and transcriptions (Kirouac, et al., 2013). The epidermal growth factor (EGF) binds to the cell-surface receptor ErbB2 and hence stimulates the phosphorylation of ErbB2. The phosphorylated receptor pERBB2 then stimulates the RAS/MEK/ERK mitogen-activated protein kinase (MAPK) cascade, activating many transcription factors correlated with tumorigenesis (Kirouac, et al., 2013; Wang, et al., 2009; Xu, et al., 2015) (Figure 4). Therefore, the ERBB-MAPK pathway is regarded as one potential target for treating cancer. MAPK inhibits the transcription factors FOXO and activates MYC, while MYC in turn functions to suppress the expressions of MAPK. The negative feedback reinforces the robustness of the regulatory network (Figure 4).

In addition, several genes involved in the TGF-β-induced pathway are differentially expressed in tumor and normal cells. The TGF-β family plays pivotal roles in regulating multiple cellular activities to regulate growth inhibition and apoptosis (Labibi, et al., 2020; Schuster and Krieglstein, 2002). TGF-β ligand binds to the receptor (TGFβR) to phosphorylate and activate the transcription factors SMAD2/3, which forms complexes with SMAD4 to control myriad downstream transcriptions, including inhibiting the activity of MYC involved in the aforementioned ERBB-regulated pathways (Gomis, et al., 2006; Labibi, et al., 2020) (Figure 4). The cross talk between different pathways combined with feedback loops contributes to the robustness of tumor growth and the resistance of anti-tumor treatments.

We proposed a mathematical model to investigate the therapeutic potential of DEX. DEX has been implicated in the suppression of ERBB-amplified proliferation pathway via disturbing the activation of MAPK (Jang, et al., 2013; Kumamaru, et al., 2011; Spinetti, et al., 2003). The activation of MAPK pathway has been verified to rescue the DEX-induced cell death (Spinetti, et al., 2003). Therefore, MAPK functions downstream in the DEX-induced apoptosis. Here, we assumed DEX inhibits the binding between ERBB and its receptor to disturb the downstream kinase cascade (Figure 4, left dashed line). Additionally, DEX activates the TGFβ-induced pathway to inhibit the expression of oncogene MYC (Massagué and Gomis, 2006; Schuster and Krieglstein, 2002). DEX causes increased protein levels of TGF-β1 and phosphorylated SMAD2 in A549 cells (Feng, et al., 2018). We found the gene expression level of TGFβR1 elevates with DEX treated (Figure 3A). Taken together, DEX likely functions upstream in TGF-β -induced pathway though the underlying mechanism is unclear. We assumed DEX amplifies the expressions of both TGF-β and its receptor (Figure 4, right dashed lines). In this study, we integrated the ERBB-amplified tumor pathway with the TGF-β-induced signaling pathway to investigate the dynamic effect of DEX in cancer (Figure 4).

By converting the regulatory network (Figure 4) into a system of ODEs (Table 1), we proposed to depict the gene regulatory relationship among A549 cells and glean insights about the effectiveness of different doses of DEX treatment for anti-tumor. Table 2 lists the model parameters and the corresponding values optimized via Genetic Algorithms (Mitchell, 1996) and Direct Search(Kolda, et al., 2003). Both the simulation and optimization were implemented in MATLAB (Fu, et al., 2020; Fu and Zhang, 2017). Ineffective empirical results obtained from the scRNA-seq dataset are ignored at this stage.

### 3.3 Biological interpretation and prediction

Our model accurately simulated gene regulatory patterns and drug response after applying DEX treatment to fully developed carcinomatous A549 cells. Based on the signature genes regulatory process, we formulated a set of ODEs to model the ERBB-amplified tumor pathway with TGF-β-induced signaling, as shown in Table 1 (Equations 1-11). As long as the DEX treatment requires a constant gene state, 1000 hours are generated for signature genes to level off before applying the drug. Following on the simulated pathways, drug and drug-associated complexes are then determined by Equations 12-13. Due to the lack of valid empirical data of EGF, ERBB2, MAPK1, and MYC, we evaluate our simulations by assessing differences of experimental data of the drug response of pSMAD2/3, TGFβ, FOXO3, and TGFβR1 with their empirical data - protein levels of SMAD2 and TGFβ1 (Feng, et al., 2018) as well as gene expression levels of FOXO3 and TGFβR1. The numerical values of experimental observations are extracted from the dataset of A549 and shown as the red circles in Figure 5.

The simulated drug response of pSMAD2/3, TGFβ, FOXO3, and TGFβR1 of our modeling precisely matches the empirical data concerning one dose of DEX treatment well. Hence, our simulation correctly captured the drug response of these signature genes. To be more specific:

**pSMAD2/3:** SMAD2/3 are downstream genes of TGF-β signaling pathways, which can translocate into the nucleus to regulate transcriptions (Labibi, et al., 2020). When DEX treatment is applied to, the phosphorylation process of SMAD2/3 is activated to reduce the protein level of oncogene. The drug response of pSMAD2/3 corresponds to the trend and fits the empirical data well, for which merely has a 0.05 shortage

**Table 2.** Parameters values

| Parameter | Value ($hour^{-1}$) | Description |
|---|---|---|
| $ks_{ef}$ | 7.3434 | Synthesis rate of EGF |
| $kb_{efeb}$ | 5.8913 | Rate of binding between EGF and ERBB2 |
| $ku_{efeb}$ | 9.5653 | Rate of unbinding of EGF-ERBB2 complex |
| $ks_{eb}$ | 3.9093 | Synthesis rate of ERBB2 |
| $ks_{mk}$ | 1.6673 | Synthesis rate of MAPK1 |
| $ks_{fo}$ | 0.6638 | Synthesis rate of FOXO3 |
| $ks_{tb}$ | 8.4325 | Synthesis rate of TGFβ2 |
| $kb_{tbtr}$ | 8.8745 | Rate of binding between TGFβ2 and TGFβR1 |
| $ku_{tbtr}$ | 8.6590 | Rate of unbinding of TGFβ2-TGFβR1 complex |
| $kb_{dgeb}$ | 7.5729 | Rate of binding between DEX and ERBB2 |
| $ku_{dgeb}$ | 2.4692 | Rate of unbinding of DEX-ERBB2 complex |
| $ks_{tr}$ | 3.4431 | Synthesis rate of TGFβR1 |
| $ks_{sd}$ | 3.1049 | Synthesis rate of SMAD2/3 |
| $ks_{mc}$ | 0.9192 | Synthesis rate of MYC |
| $kpho_{sd}$ | 0.3535 | Phosphorylation rate of SMAD2/3 |
| $kdepho_{sd}$ | 9.6013 | Dephosphorylation rate of SMAD2/3 |
| $kd_{dg}$ | $log(2)/24$ | Degradation rate of DEX |
| $kd_{ef}, kd_{eb}, kd_{efeb},$ $kd_{mk}, kd_{fo}, kd_{tb},$ $kd_{tr}, kd_{tbtr}, kd_{sd},$ $kd_{mc}$ | 1 | Rate of degradation |

| Parameter | Dimensionless | Description |
|---|---|---|
| $\alpha_{tbtr}$ | 1.6681 | Activation coefficient of TGFβ2-TGFβR1 on the phosphorylation of SMAD2/3 |
| $\alpha_{tbdg}$ | 0.2660 | Activation coefficient of DEX on TGFβ2 |
| $\alpha_{mkmc}$ | 4.9890 | Activation coefficient of MAPK1 on MYC |
| $\alpha_{trdg}$ | 0.2824 | Activation coefficient of DEX on TGFBR1 |
| $\alpha_{efeb}$ | 2.8630 | Activation coefficient of EGF-ERBB2 on MAPK1 |
| $\beta_{mkfo}$ | 9.3507 | Inhibition coefficient of MAPK1 on FOXO3 |
| $\beta_{mcmk}$ | 4.6547 | Inhibition coefficient of EGF-ERBB2 on MAPK1 |
| $\beta_{sdmc}$ | 8.3417 | Inhibition coefficient of PSMAD2/3 on MYC |
| $km_{mcmk}$ | 1.8342 | Inhibition rate of EGF-ERBB2 on MAPK1 |
| $km_{mkfo}$ | 3.4700 | Inhibition rate of MAPK1 on FOXO3 |
| $km_{psdmc}$ | 6.6297 | Inhibition rate of PSMAD2/3 on MYC |

from the empirical protein level 1.18 at 48 hours after applying DEX as shown in Figure 5A.

dynamics of variable genes in lung cancer cells following DEX treatment. With the identified key molecular markers that were associated with drug

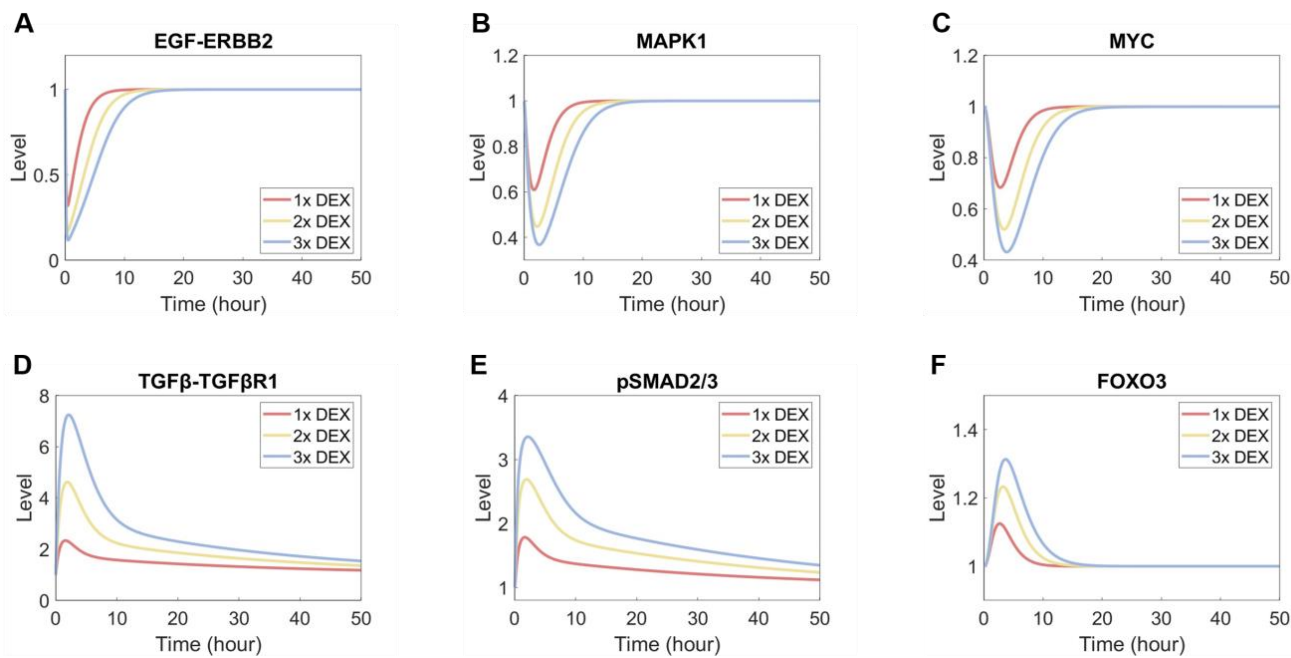

**Fig. 6. (A-F)** Model prediction of the response of EGF-ERBB2, MAPK1, MYC, TGFβ-TGFβR1 complex, pSMAD2/3, and FOXO3 after doubling (2x) and tripling (3x) the dose of DEX treatment.

**TGFβ:** In our modeling, TGFβ is activated by the treatment of DEX, and it exhibits a proper augmentation (Figure 5B) that is accord with the empirical protein level 48 hours after applying DEX.

**FOXO3:** The simulated expression level of FOXO3 locates within the three empirical values after 0, 1 and 3 hours of DEX treatment (Figure 5C). The simulated values at 1 and 3 hours immaculately match the empirical results. Nevertheless, the empirical expression level (1.117) after 2 hours relative to that after 1 hour is slightly greater than the simulated value (1.114). The minute deficit of our simulated value (0.003) may ascribe to the delayed reduction of MAPK1.

**TGFβR1:** The simulated expression level of TGFβR1 scarcely have a diversion from the empirical data, and as a DEX activating gene, it shows a proper escalating trend rightly after the application of DEX treatment as we expected (Figure 5D).

The half-life of DEX is set as 24 hours, and our simulation conforms to the drug level over time (Figure 5E). Moreover, the modeling we built looks convincing when predicting the drug response of the signature genes we mentioned under different doses of DEX treatment. As shown in Figure 6, all six genes or complexes have demonstrated an increase in protein or expression level in proportion to the increment of dose of DEX. Besides, the drug takes effect and exhausts at the same time for different doses of DEX as we could observe in Figure 6 that all three lines in different colors diverge and converge at nearly the same time.

## 4 Discussion

Though current cancer treatment had shown unprecedented success (Fu, et al., 2021; Huang, et al., 2021), the underlying mechanisms remained incompletely understood. In the present study, we utilized scRNA-seq data with bioinformatics analysis to precisely characterize the temporal response and clinical prognosis, we then depicted the dynamic signaling pathways through mathematical modeling using both scRNA-seq and proteomics data.

Our approach was essential for predicting cancer treatment response and allowing us to better understand the fundamental mechanisms of treatment. More importantly, our study provided an innovative and cross-disciplinary approach that could be further applied to immunotherapies and other cancer treatment, providing insights into the underlying mechanisms for improving cancer therapeutics, as well as implicating potential molecular targets for driving cancer resistance.

With the scRNA-seq data at different time points after treatment, we identified the variable genes with clinical significance, accompanied with knowledge support to construct the modeling components (Fu, et al., 2021). We did acknowledge our identified underlying mechanisms were not the only ones underlie treatment. Myriad crosstalk, feedback regulations, and downstream response regulators collaborate to constitute the complex regulatory network of tumorigenesis and apoptosis. However, due to the limited number of genes in this scRNA-seq data, our identified pathways were shown with the most important clues in this context(Fu, et al., 2019; Fu, et al., 2019; Fu, et al., 2021). For example, we filtered the most variable genes along the time-series data and identified the top enriched pathway compatible with our model. Additionally, though the involved genes showed significant prognostic values in discerning patient subgroups with different survival, they didn't present strong AUCs in predicting patients' survival. It is acceptable since the TCGA data is based on bulk gene expression but not single-cell level.

The comparison between experimental observation and mathematical modeling using both scRNA-seq data and proteomics data further convinces the significance of scRNA-seq data identified genes. Additionally, model prediction provides insights into potential effects of different DEX doses in cancer cells, which can be applied to investigate the dynamics and effects of downstream regulators for tumorigenesis in future studies.

## Acknowledgements

The authors acknowledge the DEMON high performance computing (HPC) cluster, the Texas Advanced Computing Center (TACC) at The University of Texas at Austin (http://www.tacc.utexas.edu), and the Extreme Science and Engineering Discovery Environment (XSEDE, which is supported by National Science Foundation grant number ACI-1548562) for providing HPC resources that have contributed to the research results reported within this paper.

## Funding

QS is supported in part by the Bioinformatics Shared Resources under the NCI Cancer Center Support Grant to the Comprehensive Cancer Center of Wake Forest University Health Sciences (P30CA012197). JS is supported by the Indiana University Precision Health Initiative.

*Conflict of Interest:* none declared.

## References

Athanasiadis, E.I., *et al.* Single-cell RNA-sequencing uncovers transcriptional states and fate decisions in haematopoiesis. *Nat Commun* 2017;8(1):2045.

Baron, M., *et al.* A single-cell transcriptomic map of the human and mouse pancreas reveals inter-cell and intra-cell population structure. *Cell systems* 2016;3(4):346-360. e344.

Bian, S., *et al.* Single-cell multiomics sequencing and analyses of human colorectal cancer. *Science* 2018;362(6418):1060-1063.

Butler, A., *et al.* Integrating single-cell transcriptomic data across different conditions, technologies, and species. *Nature Biotechnology* 2018;36(5):411-420.

Cao, J., *et al.* Joint profiling of chromatin accessibility and gene expression in thousands of single cells. *Science* 2018;361(6409):1380-1385.

Chu, L.F., *et al.* Single-cell RNA-seq reveals novel regulators of human embryonic stem cell differentiation to definitive endoderm. *Genome Biol* 2016;17(1):173.

Fabregat, A., *et al.* The reactome pathway knowledgebase. *Nucleic acids research* 2018;46(D1):D649-D655.

Feng, X.L., Fei, H.Z. and Hu, L. Dexamethasone induced apoptosis of A549 cells via the TGF-β1/Smad2 pathway. *Oncology letters* 2018;15(3):2801-2806.

Fu, T., *et al.* Pearl: Prototype learning via rule learning. In, *Proceedings of the 10th ACM International Conference on Bioinformatics, Computational Biology and Health Informatics*. 2019. p. 223-232.

Fu, T., Xiao, C. and Sun, J. Core: Automatic molecule optimization using copy & refine strategy. In, *Proceedings of the AAAI Conference on Artificial Intelligence*. 2020. p. 638-645.

Fu, T. and Zhang, Z. CPSG-MCMC: Clustering-based preprocessing method for stochastic gradient MCMC. In, *Artificial Intelligence and Statistics*. PMLR; 2017. p. 841-850.

Gomis, R.R., *et al.* C/EBPβ at the core of the TGFβ cytostatic response and its evasion in metastatic breast cancer cells. *Cancer cell* 2006;10(3):203-214.

Herr, I. and Pfitzenmaier, J. Glucocorticoid use in prostate cancer and other solid tumours: implications for effectiveness of cytotoxic treatment and metastases. *The lancet oncology* 2006;7(5):425-430.

Huang, K., *et al.* Therapeutics data Commons: machine learning datasets and tasks for therapeutics. *arXiv preprint arXiv:2102.09548* 2021.

Huynh-Thu, V.A., *et al.* Inferring regulatory networks from expression data using tree-based methods. *PloS one* 2010;5(9):e12776.

Jang, Y.-H., *et al.* Effects of dexamethasone on the TGF-β 1-induced epithelial-to-mesenchymal transition in human peritoneal mesothelial cells. *Laboratory investigation* 2013;93(2):194-206.

Jia, G., *et al.* Single cell RNA-seq and ATAC-seq analysis of cardiac progenitor cell transition states and lineage settlement. *Nature communications* 2018;9(1):1-17.

Kang, Y., *et al.* Breast cancer bone metastasis mediated by the Smad tumor suppressor pathway. *Proceedings of the National Academy of Sciences* 2005;102(39):13909-13914.

Kirouac, D.C., *et al.* Computational modeling of ERBB2-amplified breast cancer identifies combined ErbB2/3 blockade as superior to the combination of MEK and AKT inhibitors. *Science signaling* 2013;6(288):ra68-ra68.

Kolda, T.G., Lewis, R.M. and Torczon, V. Optimization by direct search: New perspectives on some classical and modern methods. *SIAM review* 2003;45(3):385-482.

Kumamaru, E., *et al.* Glucocorticoid suppresses BDNF-stimulated MAPK/ERK pathway via inhibiting interaction of Shp2 with TrkB. *FEBS letters* 2011;585(20):3224-3228.

Labibi, B., *et al.* Modeling the Control of TGF-β/Smad Nuclear Accumulation by the Hippo Pathway Effectors, Taz/Yap. *Iscience* 2020;23(8):101416.

Liu, Z., *et al.* Single-cell transcriptomics reconstructs fate conversion from fibroblast to cardiomyocyte. *Nature* 2017;551(7678):100-104.

Love, M.I., Huber, W. and Anders, S. Moderated estimation of fold change and dispersion for RNA-seq data with DESeq2. *Genome biology* 2014;15(12):1-21.

Macaulay, I.C., *et al.* Single-Cell RNA-Sequencing Reveals a Continuous Spectrum of Differentiation in Hematopoietic Cells. *Cell Rep* 2016;14(4):966-977.

Massagué, J. and Gomis, R.R. The logic of TGFβ signaling. *FEBS letters* 2006;580(12):2811-2820.

Mitchell, M. An introduction to genetic algorithms. Cambridge. In.: MA: MIT Press; 1996.

Nativio, R., *et al.* An integrated multi-omics approach identifies epigenetic alterations associated with Alzheimer's disease. *Nature genetics* 2020;52(10):1024-1035.

Patel, A.P., *et al.* Single-cell RNA-seq highlights intratumoral heterogeneity in primary glioblastoma. *Science* 2014;344(6190):1396-1401.

Schuster, N. and Krieglstein, K. Mechanisms of TGF-β-mediated apoptosis. *Cell and tissue research* 2002;307(1):1-14.

Shin, J., *et al.* Single-Cell RNA-Seq with Waterfall Reveals Molecular Cascades underlying Adult Neurogenesis. *Cell Stem Cell* 2015;17(3):360-372.

Siegel, R.L., *et al.* Cancer statistics, 2021. *CA: a cancer journal for clinicians* 2021;71(1):7-33.

Fu, T., *et al.* Ddl: Deep dictionary learning for predictive phenotyping. In, *IJCAI: proceedings of the conference*. NIH Public Access; 2019. p. 5857.

Fu, T., *et al.* Probabilistic and Dynamic Molecule-Disease Interaction Modeling for Drug Discovery. In, *Proceedings of the 27th ACM SIGKDD Conference on Knowledge Discovery & Data Mining*. 2021. p. 404-414.

Spinetti, G., *et al.* The chemokine receptor CCR8 mediates rescue from dexamethasone-induced apoptosis via an ERK-dependent pathway. *Journal of leukocyte biology* 2003;73(1):201-207.

Wang, C.C., Cirit, M. and Haugh, J.M. PI3K-dependent cross-talk interactions converge with Ras as quantifiable inputs integrated by Erk. *Molecular systems biology* 2009;5(1):246.

Wang, H., *et al.* Dexamethasone as a chemosensitizer for breast cancer chemotherapy: potentiation of the antitumor activity of adriamycin, modulation of cytokine expression, and pharmacokinetics. *International journal of oncology* 2007;30(4):947-953.

Fu, T., *et al.* HINT: Hierarchical Interaction Network for Trial Outcome Prediction Leveraging Web Data. *arXiv preprint arXiv:2102.04252* 2021.

Fu, T., *et al.* MOLER: Incorporate Molecule-Level Reward to Enhance Deep Generative Model for Molecule Optimization. *IEEE Transactions on Knowledge and Data Engineering* 2021.

Welch, J.D., *et al.* Single-cell multi-omic integration compares and contrasts features of brain cell identity. *Cell* 2019;177(7):1873-1887. e1817.

Xu, C., *et al.* Let-7a regulates mammosphere formation capacity through Ras/NF-κB and Ras/MAPK/ERK pathway in breast cancer stem cells. *Cell Cycle* 2015;14(11):1686-1697.

Yu, G.C., *et al.* clusterProfiler: an R Package for Comparing Biological Themes Among Gene Clusters. *Omics* 2012;16(5):284-287.

Zi, Z., *et al.* Quantitative analysis of transient and sustained transforming growth factor-β signaling dynamics. *Molecular systems biology* 2011;7(1):492.